\documentclass[9pt,letterpaper]{sig-alternate-05-2015}
\usepackage{layout}
\usepackage[utf8]{inputenc}
\usepackage{graphics}
\usepackage{graphicx}
\usepackage{cite}
\usepackage{xcolor}
\usepackage[]{algorithm2e}
\usepackage{listings}
\usepackage{float}
\usepackage[bookmarks=true,%
            bookmarksnumbered=true,%
            colorlinks=true,%
            linkcolor=linkcol,%
            citecolor=citecol,%
            urlcolor=urlcol,%
            hypertexnames=true,%
            pdfpagelabels]%
      {hyperref}
   \hypersetup{ pdfauthor = {gianluca stringhini},
                pdftitle = {},
                pdfkeywords = {},
                pdfcreator = {},
                pdfproducer = {}
              }

   \definecolor{linkcol}{rgb}{0,0,0.5}
   \definecolor{citecol}{rgb}{0,0.5,0.3}
   \definecolor{urlcol}{rgb}{0.3,0,0}
\usepackage{url}
\usepackage{amsfonts,pifont}
\usepackage{amstext}
\usepackage[font=footnotesize]{subfig}
\usepackage{fixltx2e}
\usepackage[keeplastbox]{flushend}
\usepackage{shortcuts}
\usepackage{booktabs}
\usepackage{enumitem} 
\usepackage{subfig}
\usepackage{array,multirow}
\usepackage{xcolor}
\usepackage{bm}
\usepackage{pgfplots}
\usepackage{xspace}
\usepackage{pifont}
\usepackage{todonotes}
\usetikzlibrary{patterns}

\newcommand{\cmark}{\ding{51}}
\newcommand{\xmark}{\ding{55}}

\begin{document}

\title{What's in a Name? \\Understanding Profile Name Reuse on Twitter}

\author{
  Enrico Mariconti$^{\star}$, Jeremiah Onaolapo$^{\star}$, Syed Sharique Ahmad$^\ddagger$, Nicolas Nikiforou$^{\star}$,\\ Manuel Egele$^\dagger$, Nick Nikiforakis$^\ddagger$, and Gianluca Stringhini$^{\star}$\\
  \affaddr{$^{\star}$University College London, $^\ddagger$Stony Brook University, $^\dagger$Boston University}\\
  \affaddr{\{e.mariconti,j.onaolapo,n.nikiforou,g.stringhini\}@cs.ucl.ac.uk}\\
  \affaddr{\{syahmad,nick\}@cs.stonybrook.edu}\hspace{3ex}\affaddr{$^\dagger$megele@bu.edu}
}

\def\startDate{October 10, 2015\xspace}
\def\endDate{April 12, 2016\xspace}
\def\totTweets{667,294,613\xspace}
\def\totUsers{70,803,606\xspace}
\def\numSharedProfileNames{106,935\xspace}
\def\totDoubleTweets{3,290,286\xspace}
\def\numUniqueUserids{196,200\xspace}
\def\numFirstUnique{101,244\xspace}
\def\numSecondUnique{15,911\xspace}
\def\numMultiple{79,045\xspace}
\def\percentageCommonTopics{40\%\xspace}

\def\numofTFIDFids{54,238\xspace}
\def\numofTFIDFpnames{26,678\xspace}

\def\percentProfilenamesWithInboundLinks{11.3\%\xspace}

\def\numSharedProfileNamesWithInboundLinks{12,037\xspace}
\def\numInboundLinks{20,457\xspace}
\def\numRankedInboundLinks{19,178\xspace}
\def\numUnrankedInboundLinks{1,279\xspace}

\CopyrightYear{2017}
\setcopyright{acmcopyright}
\conferenceinfo{WWW'17,}{April 2017, 2017, Perth, Australia}
\isbn{111-1-1111-1111-1/11/11}\acmPrice{\$15.00}
\doi{http://dx.doi.org/11.1111/1111111.1111111}

\newfont{\mycrnotice}{ptmr8t at 7pt}
\newfont{\myconfname}{ptmri8t at 7pt}
\let\crnotice\mycrnotice%
\let\confname\myconfname%

\maketitle

\begin{abstract}
 Users on Twitter are commonly identified by their profile names. These names are
 used when directly addressing users on Twitter, are part of their profile page
 URLs, and can become a trademark for popular accounts, with people referring to
 celebrities by their real name and their profile name, interchangeably. Twitter, however, has
 chosen to not permanently link profile names to their corresponding user accounts.
 In fact, Twitter allows users to change their profile name, and afterwards makes the
 old profile names available for other users to take.

 In this paper, we provide a large-scale study of the phenomenon of profile name
 reuse on Twitter. We show that this phenomenon is not uncommon, investigate the dynamics of profile name reuse, and
 characterize the accounts that are involved in it. We find that many of these
 accounts adopt abandoned profile names for questionable purposes, such as
 spreading malicious content, and using the profile name's popularity for search
 engine optimization. Finally, we show that this problem is not unique
 to Twitter (as other popular online social networks also release profile names) and argue
 that the risks involved with profile-name reuse
 outnumber the advantages provided by this feature.

\end{abstract}




\section{Introduction}\label{sec:introduction}
Users on Twitter are identified by their profile name, such as
\emph{@taylorswift13}. 
A user's profile name is also used to directly \emph{mention} accounts on
Twitter, as well as to identify their profile page's URL.\footnote{\url{https://twitter.com/taylorswift13}}
However, Twitter provides profile names as a mere convenience to its users.
Internally, the social network identifies accounts with unique numerical identifiers,
so-called \emph{user IDs} (e.g., the number 17919972 for Taylor Swift's Twitter
account).
While user IDs are globally unique and persistent, they are usually not observed
by end-users.
With these robust identifiers in place, Twitter allows users to change
the profile names of their accounts over time.
As the concepts of stable user IDs as well as changeable profile names are
crucial to this paper, we will use the terms user ID and account interchangeably
to refer to the persistent notion of an account as identified by its user ID. 
Furthermore, we use the term profile name to refer to the changeable name the
user chose for a given account.
There are multiple reasons why a user might want
to change their profile name, including changes of jobs and affiliations,
changes of names due to marriage, or the selection of a different nickname.
Interestingly, once a user changes her profile name, Twitter releases the old
name for
other users to adopt. The same happens if a user decides to delete her account. 
Only
in the case of an account that gets suspended due to a violation of the terms of
service,\footnote{\url{https://twitter.com/tos}} does Twitter make the profile name
unavailable for other users to adopt.

There are many reasons why a Twitter user might take a
profile name that was previously in use and then abandoned, both legitimate and
malicious. It is possible that a person innocently selects a profile name that
was previously in use, perhaps because it corresponds to a particularly common
first and last name (e.g., \emph{JohnSmith}). It is also possible, however, that
a malicious entity will select abandoned profile names on purpose, in an attempt
to leverage the residual reputation that this profile name might have. 
Miscreants trying to spread spam on Twitter might use this reputation in the
hope of attracting more followers. Another use of abandoned profile names is Blackhat Search Engine Optimization
(SEO)~\cite{malaga2008worst}. Since profile names identify the URL of the profile pages of
accounts on Twitter, once a profile name is freed there will be multiple links on
the web pointing to a non-existent page. By selecting an abandoned profile name
with many URLs already pointing to it, a malicious user can start promoting his content and
influence search results. 
Note that this is not a mere hypothetical scenario; the practice of harvesting abandoned Twitter profile names and using
them for SEO purposes has been observed in the wild.\footnote{\url{http://www.inetsolutions.org/quickly-easily-find-high-authority-expired-twitter-accounts/}}

We first introduced the concept of profile name reuse on Twitter in our
preliminary work~\cite{mariconti2016profile}. In that work, we showed that profile names are
reused in the wild, and we identified a number of accounts that adopted
abandoned profile names and used them to send spam.
In this paper we perform a more in-depth, larger-scale measurement of the
phenomenon of profile name reuse on Twitter. We show that over a period of one
year, 1\% of popular accounts with more than one million followers that appear in our datasets
changed their profile name, and this name was later taken by another account. We
provide a number of case studies in which a popular profile name was used to
ridicule the original owner, or to post malicious content. We also show that
Twitter users 
often prevent profile name hijacking by creating
placeholder accounts that immediately adopt the abandoned profile name and point users to
the new one. These case studies show that there is a concrete threat linked to
freeing abandoned profile names on Twitter. 

To understand how profile name reuse manifests at a large scale, we collected a 1\% sample of all public tweets
posted over a period of six months, between \startDate and \endDate. In total,
we identified \numSharedProfileNames profile names that have been shared by
\numUniqueUserids unique accounts. In doing so, we identify a set of profile names that were taken over by multiple accounts during the observation period. 
We identify different categories of accounts, ranging from those that seem to be acting for legitimate reasons to those that repeatedly iterate among different profile names with the purpose of spreading malicious content. We analyze the general characteristics of these accounts, together with the topics that they discuss on Twitter and the URLs that they post, finding that accounts that take abandoned profile names are more likely to post malicious content than regular Twitter accounts, as well as more likely to be suspended by Twitter.

Although this paper focuses on Twitter because of its size and popularity, the phenomenon of profile-name reuse is not necessarily unique to this social network. In fact, we discovered that two other major social networks (Tumblr and Pinterest) also allow profile name reuse. We conclude that freeing profile names after they are not used anymore is not a good design choice for a social network, as it exposes its users to security risks. We advocate for social networks to avoid this practice, or, at least, to monitor with particular attention accounts that adopt a previously-freed profile name. 

\noindent In summary, this paper makes the following contributions:
\begin{itemize}
\item We show that 1\% of popular Twitter accounts abandoned their profile name
  between 2015 and 2016 and had it taken over by a third party; we then provide
  a number of case studies in which abandoned profile names are used to spread
  malicious content or to ridicule the original owner.
\item We identify \numUniqueUserids accounts partaking in this practice, which shared
  \numSharedProfileNames profile names over a period of six months. We show that
  accounts that take over abandoned profile names are more likely than regular Twitter accounts to post
  malicious content and get suspended by Twitter.
 \item We show that Twitter is not the only social network allowing profile name
   reuse, but that Tumblr and Pinterest allow this practice too. 
   We argue that this practice should not be permitted because it enables
   malicious users to perform reputation hijacking and impersonation attacks.
\end{itemize}

\section{Data Collection} \label{sec:data}

Our dataset consists of a 1\% random sample of all public tweets posted on Twitter over a period of six months, between \startDate and \endDate. In total, the dataset contains \totTweets tweets posted by \totUsers distinct users (user IDs). Twitter's streaming API provides access to this data as a stream of JSON dictionaries that include information about the message and about the account that posted it. We call the dataset of \totUsers users $\textbf{U}$. We used the dataset $\textbf{U}$ for two purposes: to extract a set of popular profile names that were abandoned and taken by a third party, and to identify a set of regular accounts whose profile name was taken over. In the following, we describe these two datasets in detail.

\vspace{1ex}
\noindent\textbf{Popular account dataset.} To extract popular profile names that
were freed and taken over by a third party, we proceeded as follows.
First, we define a popular account as an account having at least one million followers. Therefore, we extracted all accounts in $\textbf{U}$ that had more than one million followers at the time the data was collected.
We identified 4,263 accounts that fit that criterion.
Second, we queried the Twitter API six month after data collection was completed to assess whether the corresponding user IDs were still associated with the same profile names that we observed during the data collection.
After this procedure, we obtained 42 popular profile names that were freed and taken by another account. This is approximately 1\% of all accounts with more than one million followers. We call this set of users, $\textbf{C}$.

For 146 popular accounts, the Twitter API did not respond with valid information.
In these cases, we visited the profile URL associated with the profile name, and we also queried the Twitter API for the user ID originally associated to that profile name. If the profile name was marked as suspended by Twitter and the user ID was not active anymore, we concluded that the original account was simply suspended for violating Twitter's terms of service. If, however, the user ID was still active but the profile name that it had freed was blocked, we concluded that some third party entity took it over and used it for malicious purposes (e.g., to spread malicious content), resulting in suspension by Twitter.

\vspace{1ex}
\noindent\textbf{General account dataset.}
Besides assessing popular accounts, we also analyzed the behavior of regular
accounts.
To this end, we analyzed $\textbf{U}$ to identify accounts that
engaged in profile name reuse during our observation period.
We identified \numUniqueUserids accounts that shared \numSharedProfileNames profile names during the
measurement period, posting \totDoubleTweets tweets. These accounts constitute 0.27\% of all accounts observed in our sample. We call this dataset
$\textbf{G}$.

\vspace{1ex}
\noindent\textbf{Auxiliary dataset.}
To better characterize accounts that partake in profile name reuse, we use a dataset of random Twitter accounts. This dataset provides a baseline of typical behavior of accounts on Twitter and helps to highlight differences in behavior shown by accounts involved in profile name reuse, compared to a regular account. To this end, we randomly sampled one million accounts from $\textbf{U}$. We call this dataset $\textbf{R}$.  

\noindent\textbf{Limitations.}
Although our dataset allowed us to measure the practice of profile name reuse on
Twitter, it has some limitations. 
First, we can only detect two accounts as sharing the same profile name if they both posted
tweets during the observation period, and if these tweets were both captured in
the 1\% sample that we collected. This means that our dataset is likely to make
us underestimate the phenomenon of profile name reuse. 

The limited visibility provided by our dataset can also impact the conclusions
that we draw from our analysis. While accounts that are observed switching
many profile names during the observation period are likely to be participating
in profile name reuse schemes and are potentially malicious, the fact that an
account is present in our dataset with a single profile name is not a guarantee
that the account never changed its profile name. As we will describe in detail
in Section~\ref{sec:analysis}, we took extra care in ensuring that we can infer
the profile name change history of an account, but some of these limitations
persist. 

\vspace{1ex}
\noindent\textbf{Ethics.}
Dealing with social network data raises ethical concerns. In this paper,
we only used publicly-available Twitter data, and we successfully obtained
ethics approval from the University College London ethics committee (Project ID 6521/004). To treat data
ethically, we followed the guidelines outlined by Rivers et al.~\cite{rivers2014ethical}.
In particular, we ensured not to link multiple datasets together with the goal of
further deanonymizing the users contained in them, and
we stored our data according to the UCL Data Protection Officer guidelines.

\section{Profile name reuse for popular Twitter accounts}

One of the main reasons for someone to take an abandoned profile name on Twitter
is to hijack the ``name recognition'' held by that account. This opens up the
opportunity for malicious actors to reach a larger audience and mount
impersonation attacks. To understand the reasons why people take over abandoned
Twitter profile names, we analyzed the accounts in the dataset $\textbf{C}$ in
detail. To recap, the profile names of these accounts belonged to popular
accounts with more than one million followers. These accounts then changed their
profile name, which was taken over by other accounts.
In total, we found that 42 profile names were abandoned and taken over by another user. This accounts for
approximately 1\% of all accounts that appear in our sample and have more that one million followers. 
Comparing this number with the profile names that are reused in the dataset $\textbf{G}$ presented in the previous section, we can conclude that popular accounts are four times more
likely to have their profile name reused than generic Twitter accounts. We then analyzed these accounts in more detail to find out the reasons behind
profile name reuse for popular accounts. Broadly speaking, we identified four
trends. 

\vspace{1ex}
\noindent\textbf{Profile names taken over by a third party with no malicious activity.}
As mentioned earlier, profile name reuse is not necessarily malicious. A Twitter user
could select a profile name that was previously in use by mere chance (e.g., in the
case of people with the same first and last names), or they could do that on
purpose, but without a malicious intention (e.g., a fan of a celebrity who has
the chance to take that person's old profile name). We observed nine cases in
which popular accounts changed their name and were taken over by someone else
without a clear malicious intention. These examples include the old profile name of
singer Lorde (\texttt{@lordemusic}), which is now owned by an account whose
tweets are private, and the old profile name of the TV show ``The Big Bang
Theory'' (\texttt{@BigBang\_CBS}), which is now owned by the official fan club
of the series.

\vspace{2ex}
\noindent\textbf{Profile names taken over by a third party to set up a parody account.}
The reputation that a Twitter profile name gains over the years makes it a
useful asset for people who want to discredit or just ridicule public
figures. We observed two cases were popular accounts changed their profile
name and their old names were taken over by parody accounts: Brazilian footballer
Alex De Souza (\texttt{@Alex10Combr}) and Colombian radio host
Vicky Davila (\texttt{@vickydavilafm}).

While parody is not a malicious activity and falls under free speech,
chances are that the original owners of those accounts did not intend to give
visibility to accounts portraying them in a satirical fashion. 
Allowing anyone to adopt a freed profile name, however, can make it easier for such parody  
accounts to gain popularity.
Moreover, not all parody accounts are
harmless. An example is what happened to Annaliese Nielsen, an
activist who recorded a video of her arguing with a minicab driver 
and threatening to ruin his reputation. This video generated outrage in certain online communities~\cite{hine2016longitudinal}, causing the
activist to be harassed on Twitter, to the point that she eventually deleted her
account. The freed profile name (\texttt{@tornadoliese}) was then taken over by trolls, who then set up a
parody account in which they ridicule the former owner. This example shows how allowing profile name reuse on Twitter can introduce a
new attack vector for online harassment. 

\vspace{2ex}
\noindent\textbf{Profile names taken over by a third party with clear malicious
intentions.} We identified that twenty profile names that were freed by their owners,
were subsequently used in violation of Twitter's terms of service, and were
eventually suspended. Examples of these profile names include highly visible
television outfits, such as, BBC Science News (\texttt{@bbcscitech}) and the
Entertainment Channel (\texttt{@eonline}). These high-profile incidents show
that attackers are actively using the reputation gained by popular profile names
to perform malicious activity. Unfortunately, we were not able to collect
evidence of the specific type of malicious activity performed by these accounts
before they were suspended by Twitter since the offending behavior was not part of our 1\% sample.

\vspace{2ex}
\noindent\textbf{Profile names ``protected'' by a placeholder account.} The
aforementioned case studies show examples of how malicious actors are misusing
abandoned profile names to their advantage. Unfortunately, Twitter does not
provide an easy countermeasure for users who want to change their profile name,
while not allowing anyone else to take the old one. We observe that some people
managing popular Twitter accounts understand the risks of profile-name reuse, and
overcome this problem by creating \emph{placeholder accounts} that take the old
profile name. These accounts usually do not post messages, but have a pointer in
their profile description to the new profile name of the account. This strategy
has been adopted by high-profile accounts such as Manchester City Football Club
(\texttt{@MCFC}) and singer Enrique Iglesias (\texttt{@enrique305}).

\section{Profile name reuse in general Twitter accounts} \label{sec:analysis}

In this section we analyze the accounts in our dataset $\textbf{G}$ in detail, with the goal of understanding
the reasons why profile names are reused on Twitter, and measure the modus
operandi of accounts that are switching between multiple profile names.
We start by defining different types of accounts involved in profile-name
reuse, and we continue with a detailed measurement of their characteristics and
activity. We then look for the presence of links pointing to abandoned
profile names, and investigate the possibility of using these links for SEO
purposes, and to inflate the popularity of Twitter accounts. Finally, we go
through some interesting case studies, exposing complex ecosystems of accounts
sharing profile names and posting about common topics.

\subsection{Types of accounts involved in profile name reuse}
\label{sec:typologies}

Accounts can be involved in profile-name reuse from various perspectives. There
are accounts changing their profile name and taking another one, which was never
held by another account, accounts that take an abandoned profile name upon
creation without knowing it, and accounts that systematically take abandoned
profile names for their gain (for example to hijack the reputation linked to
those profile names). We group these activities into three types of accounts as follows:

\vspace{2ex}
\noindent\textbf{First unique account.}
This type of account represents the typical behavior of a user changing their
profile name and taking another one that was never used before, or deleting
their account and consequently freeing its profile name. We consider an account
as belonging to this group if it is the first one in our dataset holding the
original profile name, and if it is the first one holding the new profile
name of their choice as well.  As we will explain later, if an account changes
its profile name to a name that was previously used, or changes its profile name
more than once during the measurement period, we consider this account as a
``multi account.'' Note that some misclassifications are still possible for this
category, in particular if the account used multiple profile names in the
past and this was not captured by our dataset. We observed \numFirstUnique
accounts (over \numUniqueUserids total accounts) belonging to this category 
(that is, first unique).

\vspace{2ex}
\noindent\textbf{Second unique account.} This type of account is one that
holds a single profile name in our dataset, and that profile name was freed by
another account. These accounts represent the cases in which an account takes a profile
name that was previously used, either by chance
or on purpose. One
possibility is that the abandoned profile name is a popular first and last name 
(e.g., \emph{@johnsmith}) and someone happens to have that same first and last
name, or that an account owner decides to delete their account and start over
again, using the previous profile name.
Due to the limitations in our dataset, it is possible that a
profile name switched between other profile names in the past and we did not
record that. To mitigate this problem, we look at the creation date of each
account. If the creation date is before the last time in which the first unique
user corresponding to that profile name tweeted, it means that the account was
already active back then, holding a different profile name that we did not
observe. In this case, we consider the account as a ``multi account.'' We observed
\numSecondUnique second unique accounts in our dataset.

\vspace{1ex}
\noindent\textbf{Multi accounts.} These are the accounts that switched among more than two profile names. They are either observed holding three or more profile names in our dataset (two profile names that were previously used by someone else), or they are identified by following the procedures explained in the previous two categories. Multi accounts represent a systematic behavior in which an account changed many profile names, potentially to hijack the reputation of those names. We observed \numMultiple accounts of this type in our dataset.

\begin{figure}[t]
\begin{minipage}{0.48\textwidth}
 \centering
     	\includegraphics[width=1\textwidth]{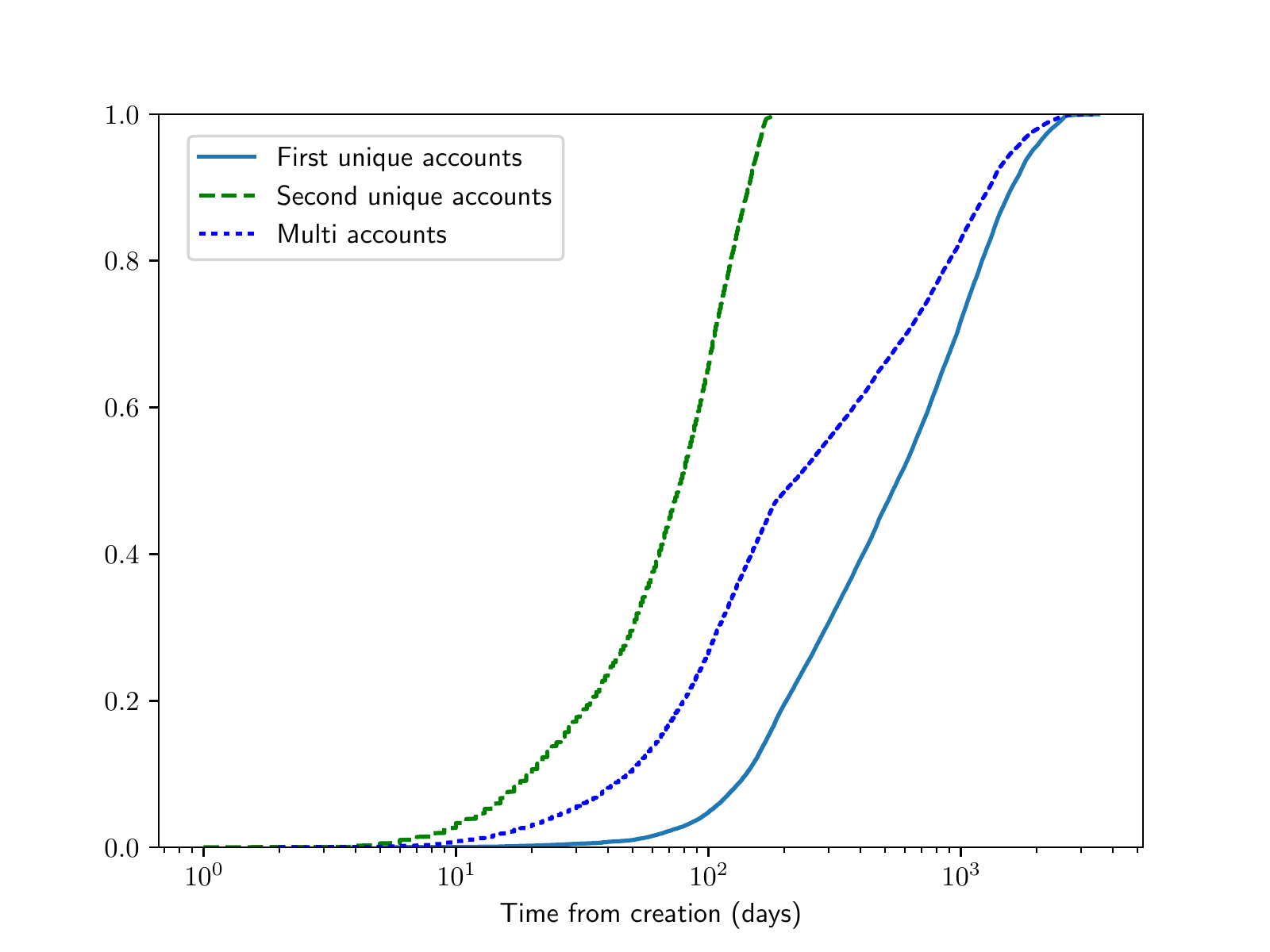}
	\caption{Cumulative distribution function of the time since the opening
	of first, second, and multi user accounts.} 
	\label{fig:cdf_time}
\end{minipage}
\end{figure}

\subsection{General characteristics of accounts reusing profile names} \label{sec:analysis_users}

\begin{figure*}[!h]
\begin{minipage}{0.48\textwidth}
 \centering
     	\includegraphics[width=1\textwidth]{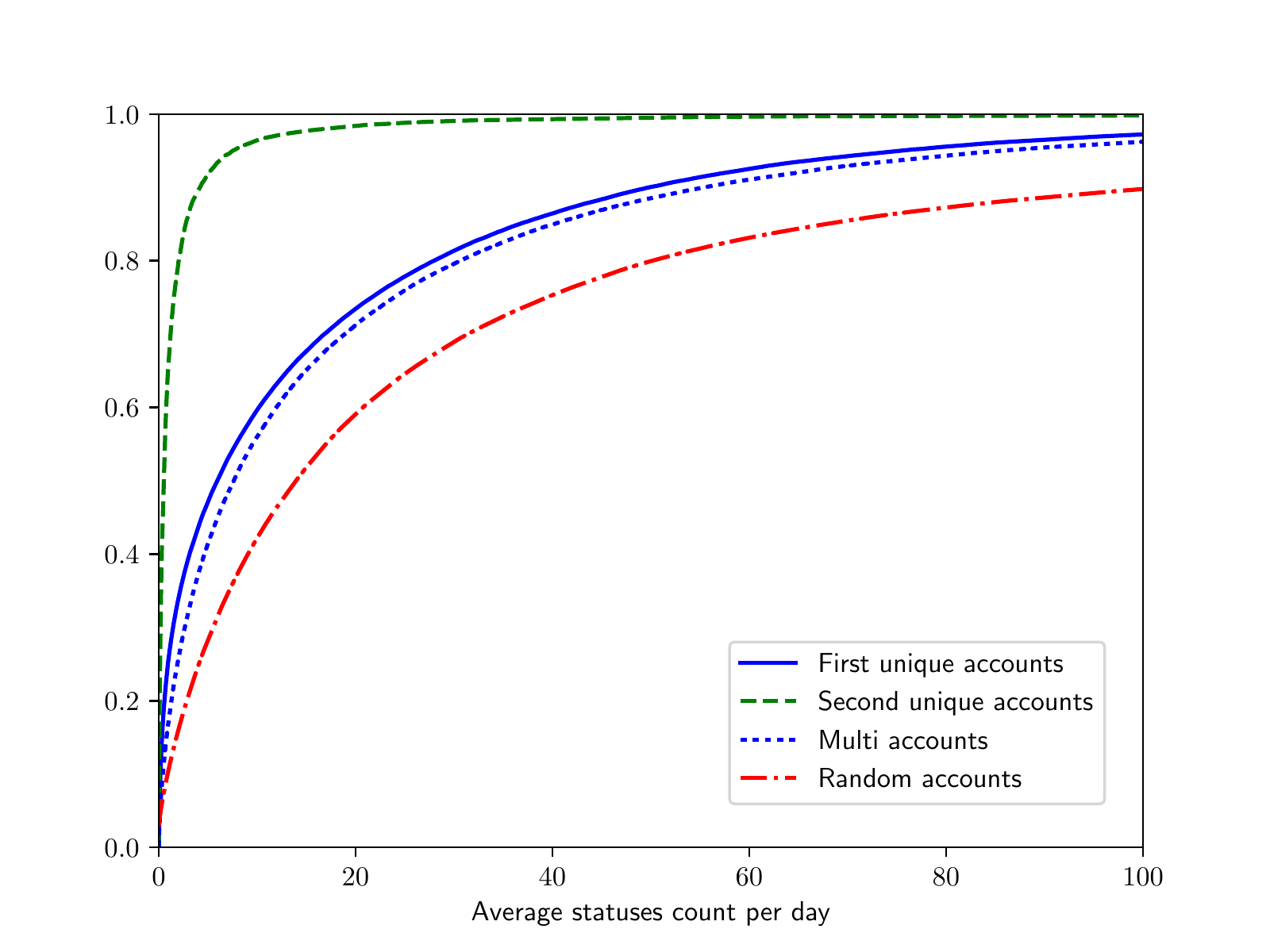}
	\caption{Cumulative distribution function of the number of tweets per day of
	first, second, and multi accounts.}
	\label{fig:cdf_statuses_average}
\end{minipage}
\hfill
\begin{minipage}{0.48\textwidth}
 \centering
     	\includegraphics[width=1\textwidth]{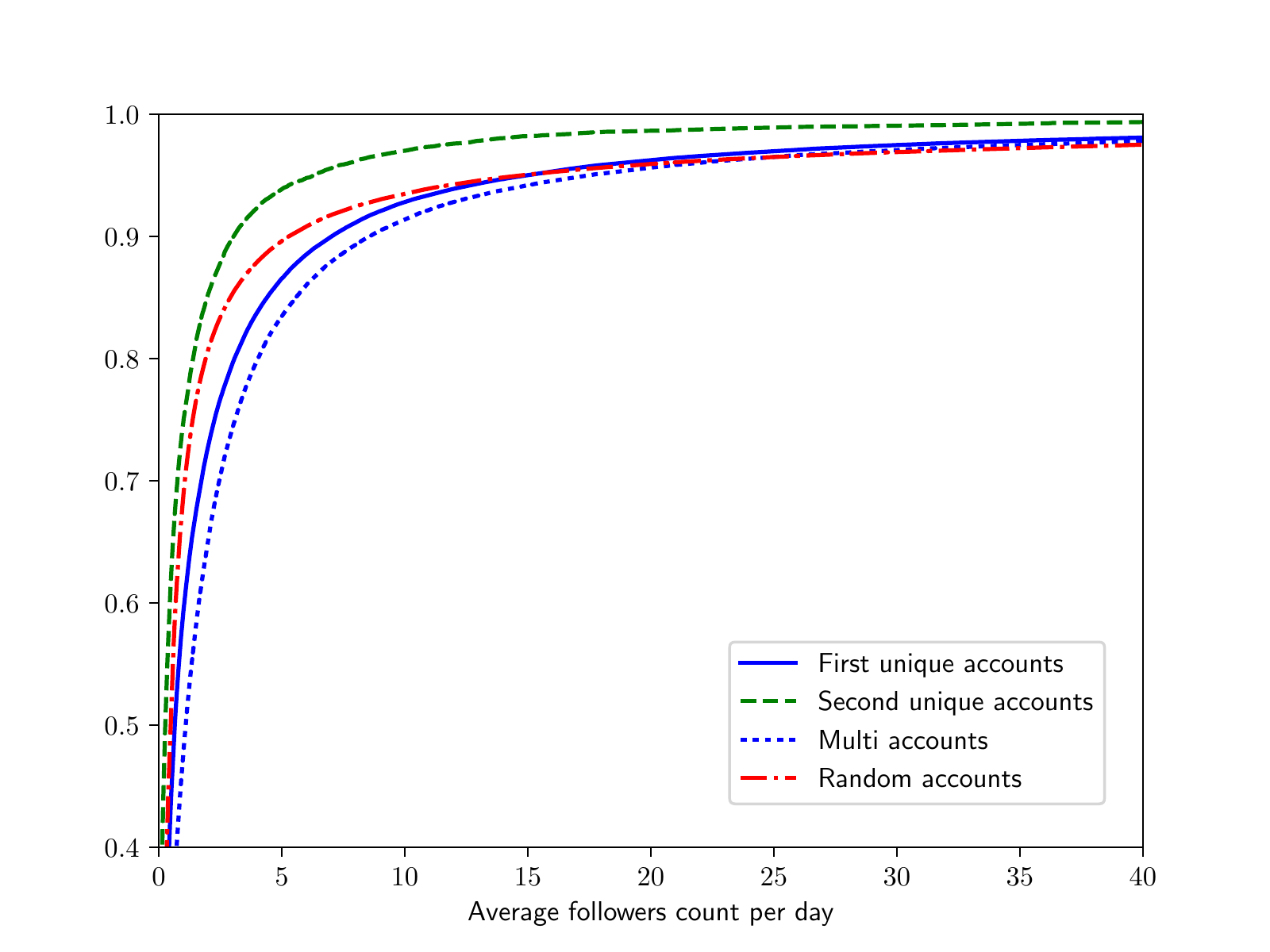}
	\caption{Cumulative distribution function of the number of followers per day of
	first, second, and multi accounts.} 
	\label{fig:cdf_followers_average}
\end{minipage}
\end{figure*}

The first analysis that we performed was looking at the general characteristics
of accounts reusing profile names. For each type of account described in the
previous section, we obtained its creation time, its number of followers, and
its number of tweets.
We first looked at the life span of accounts reusing profile names. A Cumulative
Distribution Function (CDF) of this metric for the different types of accounts
is reported in Figure~\ref{fig:cdf_time}. The figure shows that second unique
accounts are younger because, by definition, they are using only one
profile name that has already been used by someone else during the measurement
period. The earliest account creation time for them is therefore the beginning of
the measurement period. Multi accounts are also generally younger than first
unique accounts. 

We then wanted to quantify the activity levels of the different types of accounts on
Twitter. To this end, we studied the number of tweets posted on Twitter by the
various types of accounts involved in profile name reuse compared to the activity
of the random sample of one million accounts in $\textbf{R}$.
Figure~\ref{fig:cdf_statuses_average} shows the average
number of tweets posted by the accounts in our dataset per day. Accounts that
are involved in
profile name reuse generally post less tweets than the general Twitter
population. First unique and multi accounts do not seem to show a significantly
different posting behavior, while second unique accounts are much less active
than the others, with 90\% of them posting less than 10 tweets per day.
As we mentioned in Section~\ref{sec:data}, this large discrepancy between the
accounts under consideration and the random set of Twitter accounts could be due
to our collection methodology for $\textbf{R}$ accounts.

As another general characteristic of Twitter accounts, we decided to investigate the
number of followers of the accounts under consideration. The number of followers
is commonly regarded as a measure of reputation on
Twitter~\cite{stringhini2013follow}, and it ultimately regulates how many users
will see the statuses posted by an account. To exclude biases due to the age of accounts, we plotted the CDF of the average
number of followers gained per day (Figure~\ref{fig:cdf_followers_average}).
Interestingly, multi accounts are able to attract more followers on average than
general accounts, showing that switching profile name to popular ones could help in 
gaining more followers. First unique accounts also attract more followers than
random accounts, while the number of followers obtained by second unique
accounts is still less than for the other types.

\begin{figure}[t]
 \centering
     	\includegraphics[width=0.4\textwidth]{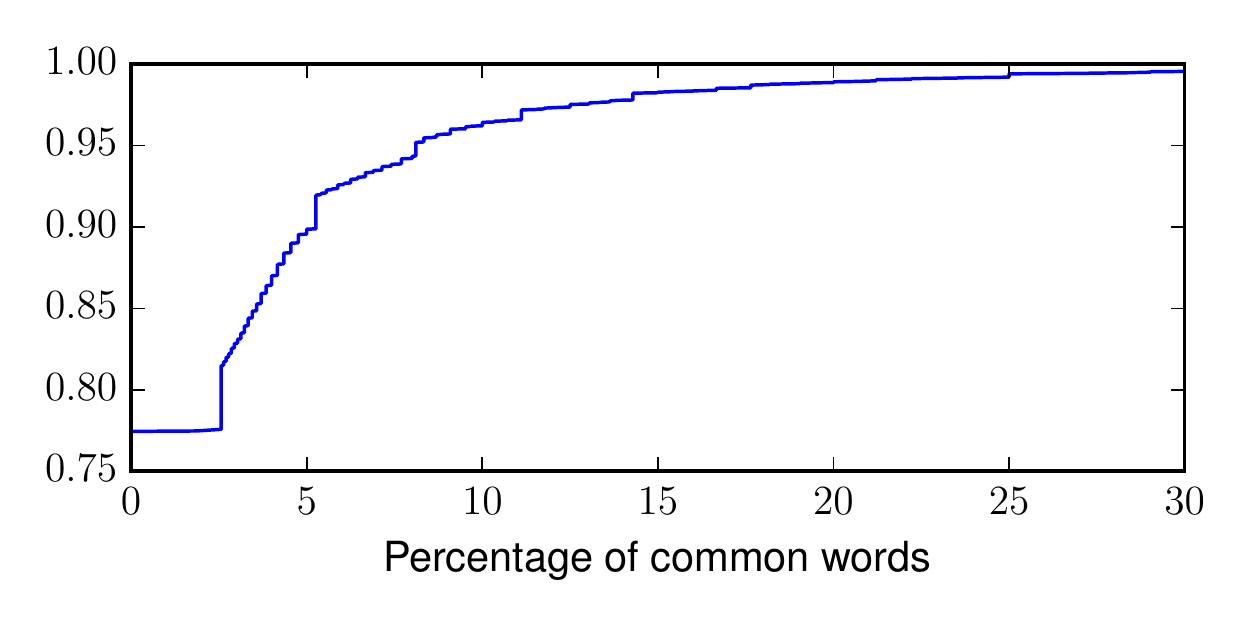}
	\caption{CDF of the percentage of common words across the tweets of the
	different accounts that shared the same profile name.}
	\label{fig:common_words_cdf}
\end{figure}

\vspace{1ex}
\noindent\textbf{Themes discussed by accounts reusing profile names.}
Next to obtaining general statistics about the activity associated with accounts involved in profile-name reuse, we wanted to understand the popular themes that these accounts tweet about. We also wanted to understand if the different accounts reusing profile names were tweeting about different themes or not. To achieve this, we conducted Term Frequency-Inverse Document Frequency (TF-IDF) analysis on the text of the tweets. TF-IDF extracts important words within a given text corpus, and the important words in turn provide information about the general theme(s) of the text corpus.

Before carrying out TF-IDF analysis, we preprocessed the text corpus as follows. First, we removed all non-English tweets; to identify such tweets, we just need to check the language from the tweet metadata contained in the JSON objects returned by the Twitter API. We could have used online automatic translation tools to translate the non-English text to English, but we found that they fail to translate some of the tweets properly. Besides, automatic translation would result in some loss of context which might bias our results. Second, we filtered out all words with less than five characters, and also removed all non-printable characters. We then carried out the TF-IDF analysis on the resulting text. In total, we processed tweets in English from \numofTFIDFids accounts associated with \numofTFIDFpnames profile names during the period of observation. 

Figure~\ref{fig:common_words_cdf} shows the CDF of the percentage of common words across the tweets of the different accounts that shared the same profile name. The words were extracted through TF-IDF analysis as earlier described. As it can be seen, 80\% of the accounts that shared profile names had less than 3\% words in common. This can be an indicator that accounts reusing a profile name are often not controlled by the previous owner, hence they tweet about completely different topics. We also grouped the important words we obtained from TF-IDF analysis into different themes. The main themes we identified included {\em Blog, Music, Porn, Video, Follow, Celebrities, Business, Retweet}, and {\em Football}. For example, the {\em Video} theme comprises the following related words: \emph{video}, \emph{youtube}, \emph{live}, \emph{stream}, and \emph{vine}. We discovered that second unique accounts are more likely to tweet about follow-back schemes, blogs, football, and business, than other themes. As we show in Section~\ref{sec:case_studies}, many second unique accounts engage in follow-back schemes. It is worth noting that, among other patterns, we discovered that a significant fraction of the tweets of multi accounts is about ``business.''

\subsection{IDs and profile-name-sharing behaviors} \label{sec:sharing}

We wanted to understand the sharing behavior of profile names by
accounts in $\textbf{G}$. In other words, we wanted to look into how many
accounts share the same profile name, and understand the complex dynamics of accounts
which change their profile name multiple times.
Most of the accounts in $\textbf{G}$ (90\%) had only one profile name. 
Similarly, most of the profile names that were involved in profile name sharing (94\%) were
using only two different profile names.
As we discussed in
Section~\ref{sec:typologies} these measurements are likely to be a lower bound of
the actual phenomenon, due to limitations in our dataset.
In Section~\ref{sec:case_studies} we will provide evidence of elaborate 
profile-name-sharing schemes, in which groups of accounts share a set of profile names among themselves.

\subsection{Suspension and deletion dynamics of\\accounts reusing profile names}
\label{ref:suspension}

\begin{figure}[t]
 \centering
     	\includegraphics[width=0.48\textwidth]{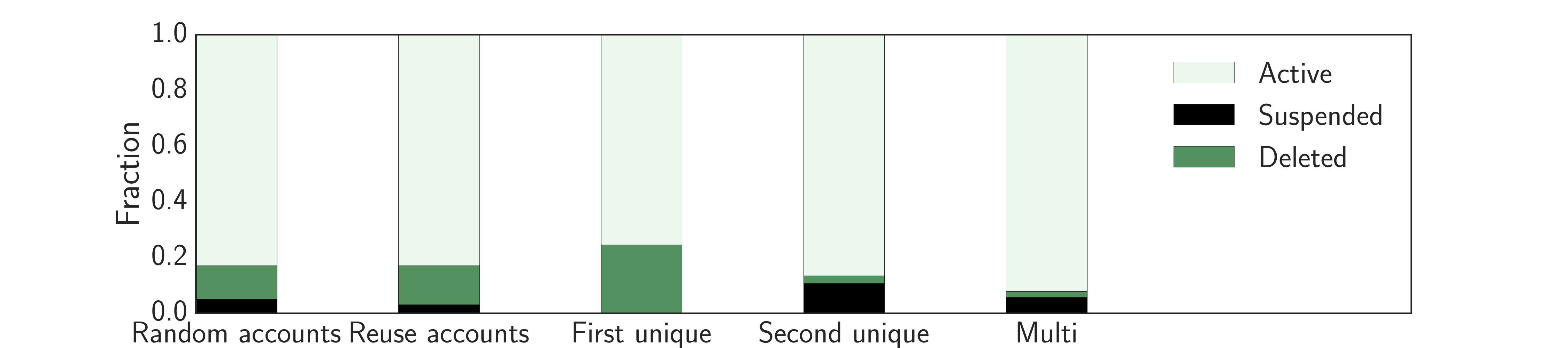}
	\caption{Ratio of user IDs sharing profile names that were suspended or deleted, compared to
	the ratio of random accounts. ``Reuse accounts'' comprises all accounts
      that shared profile names, regardless of their typology.} 
      \label{fig:suspended_user_ids}
\end{figure}

Taking another account's abandoned profile name is not an indicator of malicious
intentions per se. As we have discussed, however, the possibility of taking
abandoned profile names can facilitate malicious activity due to reputation
hijacking and impersonation opportunities. 
As an indicator of malicious activity, we wanted to understand to what extent
accounts taking part in profile name reuse get suspended by Twitter. There are
many reasons why an account can get suspended, including sending spam or
impersonating another person.\footnote{\url{https://twitter.com/tos}}

To assess the current status of a Twitter account, we proceed as follows. First, we take advantage of the Twitter API to see if an account is still active at the time of check. If it is not, it might be the case that the account was deleted by the owner, or that it was suspended by Twitter because of a violation of the terms of service. As we mentioned in Section~\ref{sec:introduction}, when an account gets suspended by Twitter its profile name does not get freed. We can take advantage of this fact to determine if the account was suspended (in case its associated profile name is not available anymore on Twitter) or whether it was deleted. This technique is not flawless, because it could happen that an account changes its profile name and gets suspended later on --- in this case, the original profile name with which we observed the account would be freed, and we would erroneously consider the account to be deleted instead of suspended. We acknowledge this problem but also believe that this happens rarely and therefore does not affect the nature of the observed trends.

We applied the aforementioned technique to our dataset of accounts involved in
profile reuse, as well as to the dataset of one million random Twitter accounts
($\textbf{R}$). Figure~\ref{fig:suspended_user_ids} reports a summary of these
results. For random Twitter accounts, the ratio of suspension is 5\%, while the
ratio of accounts being deleted is 12\%. If we look at the overall situation
of accounts reusing profile names, regardless of their type (first unique,
second unique, multi), the fraction of
suspended accounts is slightly lower (3\%) while the one of deleted accounts is
slightly higher (14\%). These numbers are however influenced by the fact that the set of
first unique accounts does not contain any suspended account. This makes sense since,
by design, first unique accounts  
released their profile name for somebody else to take,
and this profile name would have not been made available if the accounts were
suspended. 24.5\% of the first unique accounts deleted their profile during the
measurement period, while the remaining ones simply changed their profile name,
thereby releasing the old one.

Interestingly, second unique accounts have a ratio for suspension that is much
higher than random accounts (10\%). This could be due to the fact that many
of these accounts are aggressively sending spam, as we will show in
Section~\ref{sec:case_studies}.
Multi accounts have a slightly higher rate of suspension than
random accounts (5.6\%), but this small increase suggests that many of these accounts are
involved in schemes that are not clearly malicious, such as schemes for reputation boosting (see Section~\ref{sec:case_studies}).
   
\subsection{Analysis of web links towards reused\\profile names}
\label{sec:analysis_web}

In this section, we describe our analysis of the backlinks of reused Twitter user names, in an effort to quantify the misplaced trust that websites have bestowed upon these names.
To this end, we started with the set $\mathbf{G}$ of \numSharedProfileNames
Twitter profile names that were reused. We carried out this analysis immediately after the end of the data collection phase.
We used Moz-Open Site
Explorer\footnote{\url{https://moz.com/researchtools/ose}} to gather inbound
links for each Twitter profile name contained in our set. We discovered that out
of the \numSharedProfileNames profile names,
\numSharedProfileNamesWithInboundLinks (\percentProfilenamesWithInboundLinks) of
them had active inbound links on the web. The total number of discovered inbound
links for these \numSharedProfileNamesWithInboundLinks profile names was \numInboundLinks. Next, for each page hosting a link to a reused Twitter account, we gathered three additional attributes, namely, its Rank, Safety, and Category. The Rank attribute is
the Alexa rank of a linking page's main domain (TLD + 1) and was sourced from lookups in Alexa's top one million dataset, as well as scraping of Alexa's API\footnote{\url{http://data.alexa.com/data?cli=10&url=alexa.com}} for the links having ranks greater than one million. In this manner, we were able to
collect Rank information for \numRankedInboundLinks pages, leaving
\numUnrankedInboundLinks pages without rank information. 
The categories of Safety and Category were provided by Trendmicro's public website categorization engine\footnote{\url{http://global.sitesafety.trendmicro.com}}. We found that the vast majority of domains linking to the hijacked Twitter user names are safe sites and not exploit-ridden sites that link to Twitter accounts for SEO purposes.
We also found that the vast majority of websites linking to reused Twitter user names are popular benign websites with ranks that imply thousands of visitors on a daily basis.

The discovered websites belong to a wide range of categories. The top three categories are Social Networks, Computers and Internet, and Sports, in that order. The remaining categories include Entertainment, News/Media, and Politics, among others.  

\begin{figure}[t]
\begin{minipage}{0.48\textwidth}
  \begin{center}
    \includegraphics[width=1\textwidth]{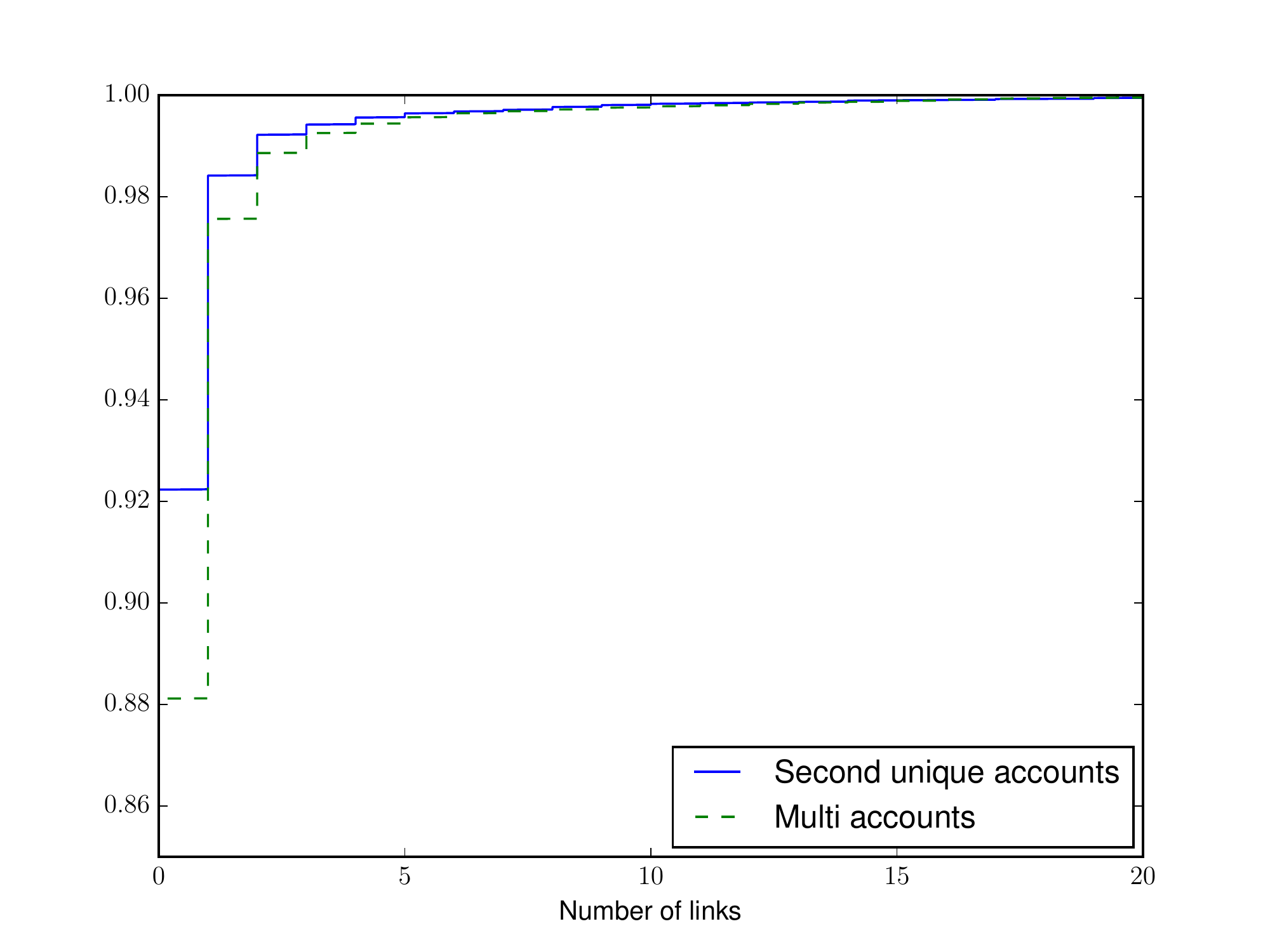}
 \caption{CDFs of link distribution per group of accounts associated with reused Twitter user name.}
 \label{fig:link_distribution}
 \end{center}
\end{minipage}
\end{figure}

We wanted to see if multi accounts have a preference for reusing profile names
with more links pointing to them. 
We counted the number of links pointing to profile names that were reused at some point during our observation, grouping the accounts as earlier described (that is, second unique and multi account groupings). We found that users in the multi accounts category have a higher preference for reusing profile names that have more links pointing to them, than second unique accounts, likely for SEO purposes. Figure~\ref{fig:link_distribution} highlights this observation, which we confirmed by applying chi-square tests to the data. Specifically, we tested the hypothesis that multi accounts have more websites linking to them with respect to second accounts. We tested the two categories by dividing the accounts in those to which at least two links in the wild were pointing at or not. The p-value of the statistical test is less than 0.0001, stating that there is evidence of difference between multi and second accounts. 

Multi accounts have more websites pointing to them with respect to second unique ones, therefore this means that multi accounts are more likely to use profile names that have multiple links on the web pointing to them. We further observed that the majority of Twitter accounts in our dataset that reuse profile names and have inbound links have less than ten such links, with around 90\% of accounts having five or less inbound links.

Overall, if we combine the information of ranking, category, and links pointing to reused profile names, we can safely conclude that the majority of websites linking to these hijacked Twitter user names are popular benign sites which have no knowledge of the fact that they are no longer linking to the accounts of popular users and celebrities, but rather to accounts that are now under the control of potentially malicious users. This confirms that profile name reuse in social networks is far from a theoretical danger since it is already happening in the wild.

\subsection{Case studies}
\label{sec:case_studies}

In this section, we analyze two case studies of accounts reusing
profile names. These examples show different ways in which abandoned profile
names are used in the wild.

\vspace{1ex}
\noindent\textbf{Activity of second unique accounts.} In
Section~\ref{ref:suspension} we showed that second unique accounts have a higher
chance to get suspended by Twitter (10\% compared to 5\% for a random population
of accounts). We investigated possible reasons why these accounts could have
been suspended. We could not identify a common theme, since the activity of most
accounts seemed unrelated to each other. We, however, identified evidence 
of accounts sharing links to YouTube videos that are now deleted for
violation of their terms of service, as well as links pointing to malware and
pornography. We also identified accounts from this type engaging in follow-back
schemes~\cite{stringhini2013follow}. Our hypothesis is that these accounts obtained abandoned profile
names and started posting malicious content, hoping to leverage the residual
popularity of these accounts to attract more victims.

\begin{figure*}[t]
 \centering
    \includegraphics[width=1\textwidth]{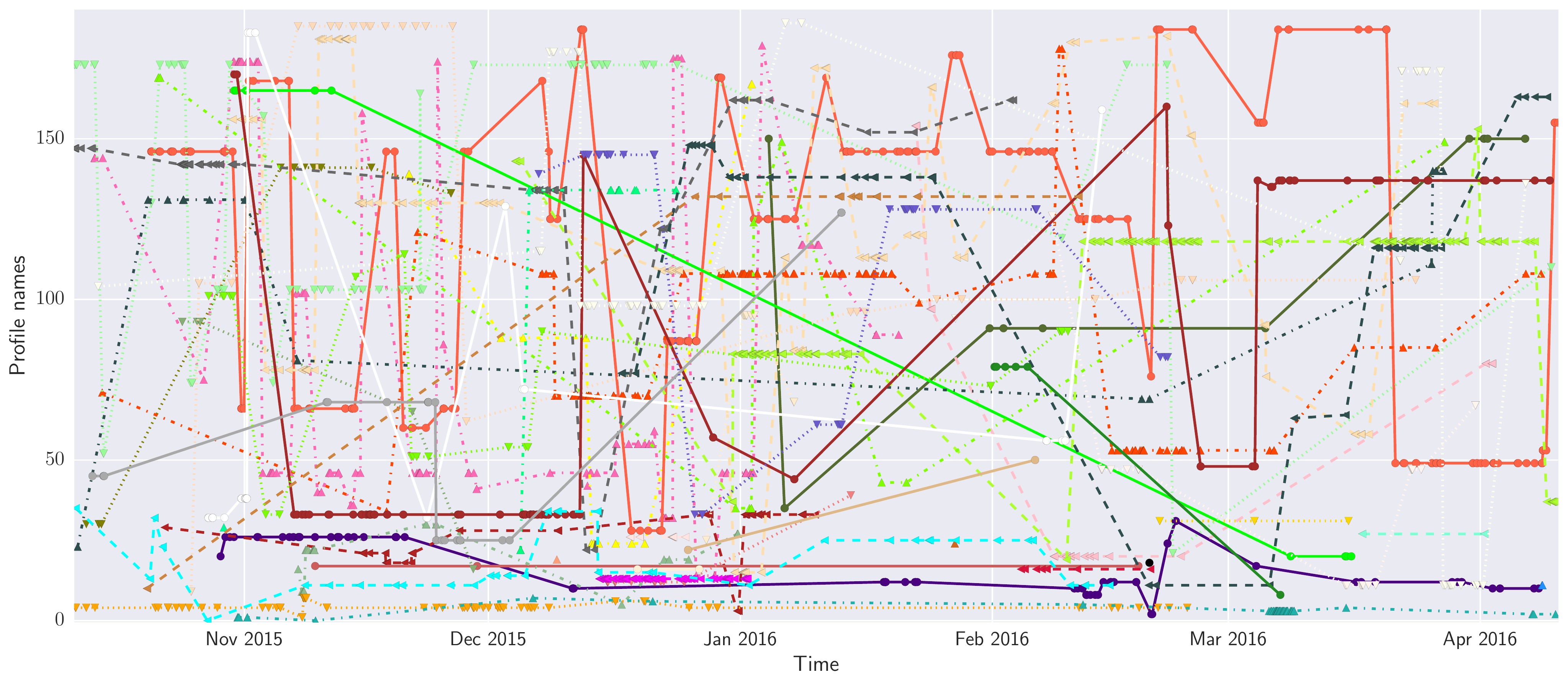}
	\caption{Timeline of a group of 48 accounts that shared 187 distinct profile names during the measurement period. Each line shows a different
      account, while the Y axis shows different profile names. Each dot represents a tweet sent by the account with a certain profile name. As it can be seen, accounts in this group typically changed profile name multiple times over the measurement period. It also happened in multiple occasions that a previously freed profile name was taken by another account in the group.}
	\label{fig:case_study2}
\end{figure*}

\vspace{1ex}
\noindent\textbf{A group of accounts sharing profile names.} We identified
an interesting group of 48 accounts that shared 187 distinct profile names. Every time one of the accounts released a profile name, another one took it. Figure~\ref{fig:case_study2} shows the timeline of this case study. Overall, the accounts involved in it changed profile name 246 times, often taking
names that were previously abandoned. 
These accounts appeared to portray models from Asia, and the accounts were tweeting in Indonesian. 
One of the popular words that appeared in the tweets sent by these accounts was ``follow,'' indicating that the accounts that reused those profile names were likely 
involved in follow-back schemes, similar to the ones presented in previous work~\cite{stringhini2013follow}.

\section{Discussion}

In this section we first aim to understand whether the problem of profile name
reuse is unique to Twitter. We then reason about our findings and offer some
suggestions that social networks should put in place to avoid the security
problems highlighted in this paper.

\subsection{Profile name reuse on other\\social networks}

\begin{table}
  \begin{center}
    \begin{tabular}{|l|r|r|}
      \hline
      \textbf{Service} & Allows change & Allows reuse \\
      \hline
      Facebook & \cmark & \xmark \\
      Google+ & \xmark & \xmark \\
      LinkedIn & \cmark & \xmark \\
      \color{red}{\textbf{Pinterest}} & \color{red}{\cmark} & \color{red}{\cmark} \\
      Reddit & \xmark & \xmark \\
      Snapchat & \xmark & \xmark \\
      \color{red}{\textbf{Tumblr}} & \color{red}{\cmark} & \color{red}{\cmark} \\
      \color{red}{\textbf{Twitter}} & \color{red}{\cmark} & \color{red}{\cmark} \\
      Youtube & \xmark & \xmark \\
      \hline
    \end{tabular}
    \caption{Possibility of changing user name on popular online social
    networks. With the exception of Google+, Youtube, Reddit, and Snapchat, all other popular social networks allow
  users to change their user name. Moreover, Twitter, Pinterest, and
Tumblr allow users to take a user name that used to belong to another account. This policy makes these social networks vulnerable to user name squatting.}
    \label{tab:change}
  \end{center}
\end{table}

In this paper, we analyzed the phenomenon of profile name reuse on Twitter.
As a further step, we wanted to understand to what extent other popular social networks allow their users to
change their user name. To this end, we examined the terms of service of eight
popular social networks other than Twitter: Facebook, Google+, LinkedIn, Pinterest, Reddit, Snapchat, Tumblr, 
and Youtube. We looked for clauses specifying whether a user is allowed to
change their user name and if the released user name can be used by other users
in the future. In case the terms of service did not contain a definite answer to
these questions, we created two accounts on the platform, $a_1$ and $a_2$. We
then changed $a_1$'s user name (if that was allowed), and set up $a_2$ to pick
$a_1$'s old user name, checking if this was allowed. 
The results of our investigation are summarized in Table~\ref{tab:change}. We
identified three categories of social networks, illustrated in detail in the
following.

\vspace{1ex}
\noindent\textbf{Social networks not allowing a change of user name.} These
social networks present the strictest settings, not allowing users to change
their user name at all. Reddit\footnote{\url{https://www.reddit.com/r/help/wiki/faq}}  and Snapchat\footnote{\url{https://support.snapchat.com/en-US/a/change-username}} belong to this group.
Google+ and Youtube allow users to change the name of their profile up to three
times in 90 days. This name is the one listed on the profile page of a user and
used to directly mention her, but is not part of the URL of the profile page. 
At the same time, users are also allowed to set a \emph{handle} for their page or
channel, which allows people to more easily remember the URL associated to it (e.g.,
\emph{https://youtube.com/taylorswiftVEVO}). After this name has been set, however,
it is not possible to change it
anymore.\footnote{\url{https://support.google.com/plus/answer/2676340}} 
We mark both Google+ and Youtube
as social networks that do not allow a profile name change. 

\vspace{1ex}
\noindent\textbf{Social networks allowing a change of user name, but not its
reuse.} Similar to Google+ and Youtube, Facebook and LinkedIn also allow users
to change their name --- a legitimate use of this being, for example, a user
changing their last name after their spouse's. Moreover,
Facebook\footnote{\url{https://www.facebook.com/help/203523569682738}} and LinkedIn\footnote{\url{https://www.linkedin.com/pulse/20140424124611-12064186-how-to-customize-your-\\linkedin-public-profile-url}} allow users to change the user name that is associated to
the URL of their public profile page.
Facebook, however, limits their users to a single change of user name in the
account's lifetime. More importantly, after a user name is changed on these networks, it
is not made available for others to use. Facebook and LinkedIn, therefore, are not
vulnerable to the type of impersonation attacks described in this paper.

\vspace{1ex}
\noindent\textbf{Social networks allowing both a change of user name and its
reuse.} Finally, Pinterest\footnote{\url{https://help.pinterest.com/en/articles/edit-your-profile}} and
Tumblr\footnote{\url{https://www.tumblr.com/docs/en/blog_management}} match Twitter's capability in profile
name reuse,  both allowing users to change their user name
\emph{and} return the old user name to the pool of available ones. A user name
on these networks identifies both the user, and the URL of their profile page (or
their blog in the case of Tumblr). As a partial mitigation, Tumblr releases the
old user name only after 24 hours from the change. 
These findings show that the security issues and the phenomena highlighted in
this paper are not unique to Twitter, but can be found on other
social networks too.

\subsection{Recommendations to social networks}
By using large-scale measurements and individual case studies, in this paper we described the problem
of profile-name reuse on Twitter and how other social networks that follow similar name-reuse choices
will likely suffer from the same type of abuse. 
We acknowledge that allowing profile name reuse can have some benefits and it is
a user friendly policy that gives a high degree of freedom to users. We showed,
however, that this policy also has security implications.

The best solution to these security issues is, in our opinion, not to allow accounts to reuse abandoned
profile names. Some social networks such as Facebook and LinkedIn allow the
change of user name, but they do not allow the reuse of profile names that have
already been used by someone; we suggest that Twitter, Tumblr, and Pinterest should adopt a similar policy to easily and effectively tackle the problem. 
As a less strict approach, social networks could allow accounts to take
abandoned profile names, but should then start monitoring them for signs of malicious activity.

\section{Related Work}

Being one of the most popular online social networks, Twitter attracted
significant interest from the research community, who studied its general
characteristics~\cite{Krishnamurthy:2008:usenix}, how reputation on the network
works~\cite{Cha:2010:aaai}, investigated peculiar traits of the service, in
particular its microblogging features~\cite{Kwak:2011:FOR}, and looked at the
unfollow patterns of Twitter users~\cite{Kwak:2010:10:www}. 

Particular focus was given to the security issues around Twitter.
Grier et al. performed the first large-scale study of abuse on the
platform~\cite{Grier:2010:ccs}, while Thomas et al. studied marketplaces where one can buy fake Twitter
accounts~\cite{thomas2013trafficking}.
Stringhini et al. studied services that sell
compromised accounts as followers to customers that are willing to pay for
them~\cite{stringhini2013follow,poultry}. De Cristofaro et al. studied the
ecosystem of services that deliver likes to the Facebook pages of their
customers~\cite{de2014paying}. 
Based on the insights from this research, a number of systems have been proposed
to detect malicious activity on Twitter, such as fake
accounts~\cite{Stringhini:10:socialnet-spam,Thomas:2011:oakland,Benvenuto:2010:CEAS,Lee:2010:SIGIR,Lee:2012:ndss}, compromised accounts~\cite{egele13:compa} or malicious accounts controlled by botnets~\cite{stringhini2015evilcohort,zhao2009botgraph,cao2014uncovering}. 
Goga et al. studied the problem of impersonation on Twitter~\cite{gogadoppelganger}. 
In this work, we showed how reusing abandoned profile names could facilitate the impersonation problem on Twitter.

The problem of profile name reuse on Twitter was originally presented in our
preliminary work~\cite{mariconti2016profile}. In that previous paper, we
identified the security issues linked to profile name reuse on Twitter,
identified 19,000 profile names that had been reused over a period of one
month, and provided general statistics about them. In this paper, we took the
study much further, analyzing more accounts for a much longer period of time,
and performing a deeper analysis on the modus operandi and characteristics of
accounts that reuse abandoned Twitter profile names.
Jain et al. also published a study in which they show that users on Twitter
temporarily change their profile names~\cite{change2016}.

The problem of profile name reuse shares certain similarities with the phenomenon of
\emph{cybersquatting} since attackers essentially squat profile names that they do not own in 
an attempt to confuse visitors about the nature of a Twitter account. In a similar fashion,
in domain squatting, attackers register domains that are confusingly similar to popular authoritative
domain names, and abuse this similarity for various types of advertising fraud, phishing, and malware
delivery~\cite{edelman2003,Wang:2006:STD:1251296.1251301,Banerjee2008,Moore2010,Szurdi:long-taile-of-typosquatting,Agten2015seven,khan2015every}.

\section{Conclusion}

In this paper we studied the phenomenon of profile name reuse on Twitter. We
identified a number of interesting ways in which profile names are reused, some
of which are malicious. We also showed that Twitter is not the only social
network vulnerable to the issues highlighted in this paper. We hope that this
work will help to raise awareness of the issues with freeing profile names after
they have been abandoned.

\section*{Acknowledgments}

We wish to thank the anonymous reviewers for their comments. This work was
funded by the H2020 RISE Marie Sklodowska Curie action (MSCA) grant number
691925, by the EPSRC grant number EP/N008448/1, by the NSF under grant
CNS-1527086, and by a Boston University Hariri Institute for Computing Research
Award. Enrico Mariconti was funded by the EPSRC under grant 1490017, while
Jeremiah Onaolapo was supported by the Petroleum Technology Development Fund
(PTDF), Nigeria.

\bibliographystyle{abbrv}
\bibliography{biblio}

\end{document}